# Forward stimulated Brillouin scattering in silicon microring resonators

Yaojing Zhang,[1] Liang Wang, [1,a)] Zhenzhou Cheng,[2] and Hon Ki Tsang[1,a)]

[1]*Department of Electronic Engineering, The Chinese University of Hong Kong, Shatin, New Territories, Hong Kong*

[2]*Department of Chemistry, The University of Tokyo, Tokyo 113-0033, Japan*

Stimulated Brillouin scattering (SBS) has been demonstrated in silicon waveguides in recent years. However, due to the weak interaction between photons and acoustic phonons in these waveguides, long interaction length is typically necessary. Here, we experimentally show that forward stimulated Brillouin scattering in a short interaction length of a 20 μm radius silicon microring resonator could give 1.2 dB peak gain at only 10mW coupled pump power. The experimental results demonstrate that both optical and acoustic modes can have efficient interaction in a short interaction length. The observed Brillouin gain varies with coupled pump power in good agreement with theoretical prediction. The work shows the potential of SBS in silicon for moving the demonstrated fiber SBS applications to the integrated silicon photonics platform.

Stimulated Brillouin Scattering (SBS) has been attracting research interest for many years. SBS is a nonlinear interaction between optical and acoustic fields[1]. It has been widely studied in optical fibers[2,3]. Many applications based on SBS have been demonstrated including SBS-based slow light for fiber-optic temperature sensor[4], SBS enhanced four-wave mixing (FWM) in optical fibers [5] and SBS in optical fibers for high resolution optical spectrum analyzer[6]. SBS has also recently been demonstrated in mm scale silicon pedestal waveguides[6,7].

The high refractive index contrast silicon-on-insulator (SOI) waveguide platform has been extensively used to study nonlinear effects in silicon waveguides[8-10] including FWM[11], self-phase modulation[12] and stimulated Raman scattering[13]. In the case of stimulated Brillouin scattering, the leakage of phonons to the silica substrate prevented observation of SBS in SOI platform until the recent introduction of novel structures to confine the phonons by a local etching of the buried oxide in a pedestal waveguide structure[6] or a fully suspended silicon membrane structure[14]. Due to the weak overlap between guided optical and acoustic modes, relatively long lengths in the mm or cm range are necessary for generation of SBS in these structures.

Cavity structures such as ring resonators have build-up of optical intensity in the cavities and generally offer devices with much smaller footprints. With the use of high Q-factor resonators, nonlinearities such as FWM have been demonstrated with low input power[15]. Recently 1.5cm round trip length hybrid $As_2S_3$ on silicon ring resonators were used to achieve stimulated Brillouin lasing with 50mW coupled pump powers[16]. Here we study the possibility of optical gain from SBS in silicon microring

___________________________

a) Electronic mail: hktsang@ee.cuhk.edu.hk, lwang@ee.cuhk.edu.hk

resonators where the pump and stokes shifted probe signals are around the same order of resonance in the microring resonator rather than the adjacent resonances used in the long ring resonator for the SBS laser[16]. Since light-sound overlap in backward SBS has been reported to be relatively small in silicon structure compared to large overlap in forward SBS[6], we focus on forward SBS in this letter.

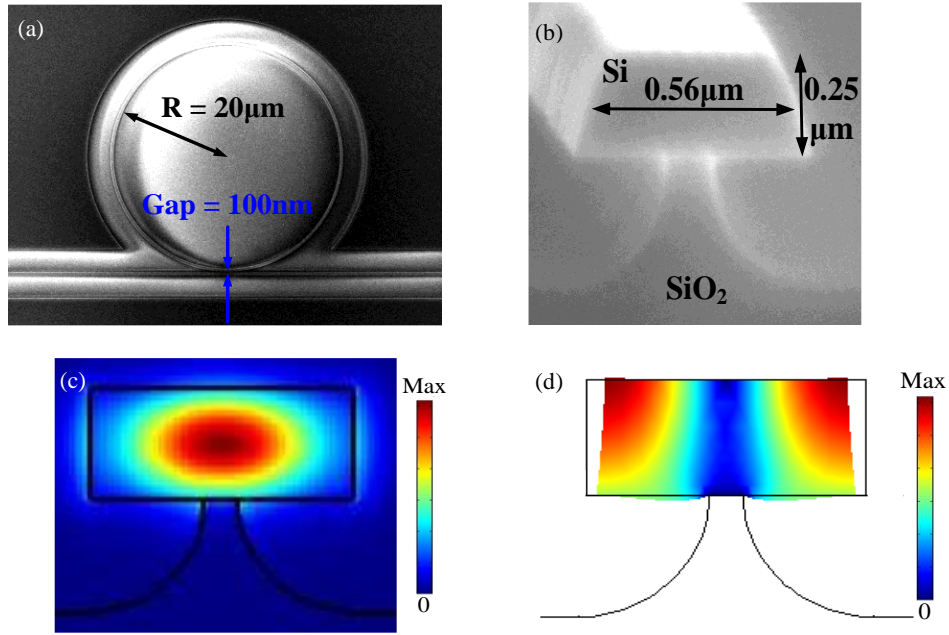

FIG. 1. Scanning electron microscope (SEM) images of (a) silicon microring resonator in the experiment and (b) its cross section which is supported by a narrow pillar. (c) The electric field profile of fundamentally optical mode and (d) the transverse profile of elastic displacement field.

The microring resonator is shown in FIG. 1 (a) and consists of a ring with radius of 20 μm and bus waveguide with length of 220 μm. The gap between the ring and the bus waveguide is 100 nm. The widths of both ring and bus waveguide are 560 nm as shown in FIG 1. (b). The resonator supports the fundamental transverse electric mode shown in FIG. 1 (c) and the acoustic phonon mode at about 7.8 GHz frequency shown in FIG. 1 (d). Even though there remains some leakage of phonon due to the supporting $SiO_2$ pedestal, the recirculating nature at resonance in the microring resonator effectively enhances the photon-phonon interaction length. We use electron-beam lithography (ELIONIX) to define the device layout on the SOI wafer (SOITEC) with 0.25 μm top silicon layer and a 3 μm buried-oxide (BOX). The waveguides were produced by dry etching to the buried oxide using $C_4F_8$ and $SF_6$ gases. The pedestal structure shown in Fig 1(b) was produced by wet etching using 10% diluted hydrofluoric acid solution at etching rate of 77 nm/min.

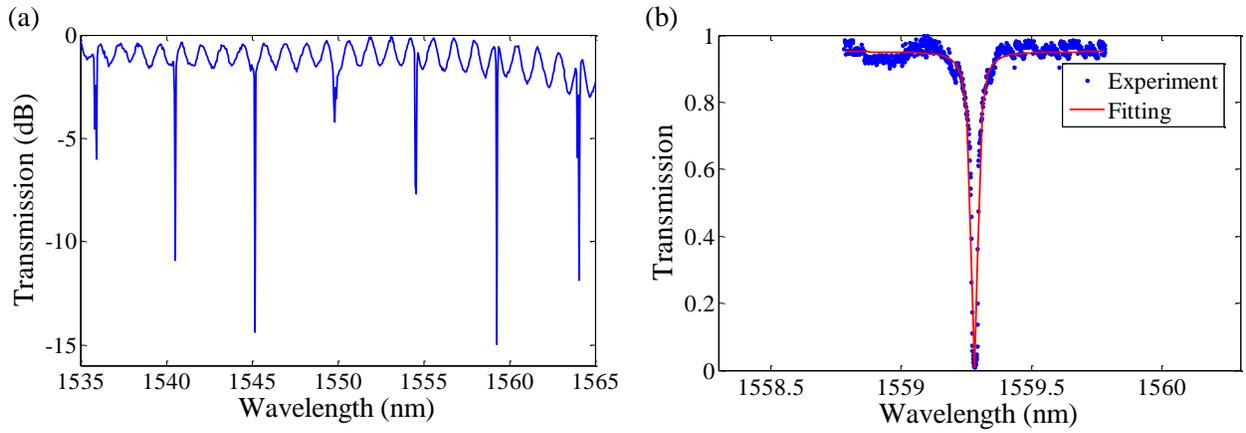

FIG. 2. Measured spectra of microring resonator. (a) Transmission spectrum of the microring resonator from 1535 nm to 1565 nm; (b) spectrum measured around 1559.28 nm resonant wavelength and corresponding fitting curve.

The spectral response of the microring resonator was obtained using a tunable laser to scan input pump wavelength from 1535 to 1565 nm and an optical spectrum analyzer (OSA) with fine resolution of 0.02 nm to record the spectrum. We obtained the transmission spectrum of the resonator as shown in FIG 2. (a). The free spectral range (FSR) of the microring resonator is measured to be about 4.5 nm at the resonant wavelength around 1559.28 nm which agrees well with the calculation of FSR = $\lambda_{res}^2/(n_g L)$ ($\lambda_{res}$ is the resonance wavelength, $n_g$ is the group index and L is the round trip length)[17]. Fine wavelength scan around 1559.28 nm with a wavelength step size of 1 pm allowed the measurement of the loaded Q-factor of the microring resonator. With Lorentzian curve fitting, the loaded Q-factor was measured to be ~62300 as shown in FIG 2. (b). Meanwhile, using

$$Q = \frac{\pi n_g L \sqrt{ra}}{\lambda_{res}(1-ra)} \qquad (1)$$

where r is the transmission coefficient of the bus waveguide, $n_g$ is the group index, L is the round trip length, $a$ is the single-pass amplitude transmission related to the round trip loss α by $a^2 = \exp(-\alpha L)$, $\lambda_{res}$ is the resonance wavelength[17], we can estimate the microring resonator parameters r and α to be 0.99 and 2 dB / cm respectively.

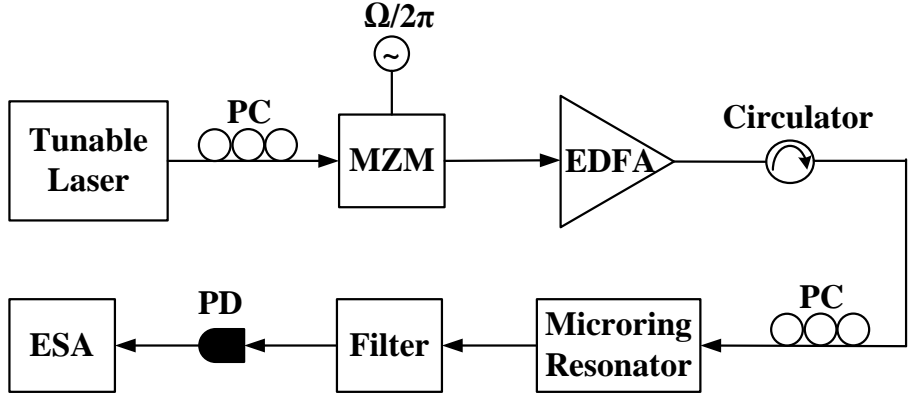

FIG. 3. Experimental setup to measure forward SBS in the microring resonator. PC: polarization controller. MZM: Mach-Zehnder modulator. EDFA: erbium-doped fiber amplifier. PD: photodetector. ESA: electrical spectrum analyzer.

We used the experimental setup shown in FIG. 3 to measure the forward SBS in the resonator. The pump laser was a tunable laser which had a linewidth of ~10 kHz (Pure Photonics). The laser wavelength was scanned at a step size of 1 pm which allows accurate approach to the resonance around 1559.28 nm. The laser output was modulated by a dual-drive Mach-Zehnder modulator to produce a redshifted sideband with frequency detuning of $\Omega$ from the pump laser and serve as the probe. The frequency detuning was varied by tuning the signal generator. Both pump and probe are amplified by an EDFA. A circulator was used to prevent any backward SBS from coupling into the EDFA. Mechanical fiber PCs adjusted the polarization of the input light, and the light was coupled to the silicon waveguide via subwavelength grating couplers. We used WaveShaper (Finisar) as an optical filter with a stopband to remove output at the anti-Stokes frequency. The output from the filter was coupled to a 20 GHz photodiode which was connected to an ESA. The amplitude at the probe frequency detuning was measured in the ESA because of the beating between the pump and probe light at the photodiode[14]. The Brillouin gain spectrum was obtained by manually sweeping $\Omega$ with 5 MHz tuning step and observing the relative change in the intensity of the probe[14]. Because the ESA display was calibrated for electrical power (proportional to the square of the photocurrent) instead of optical power (proportional to the photocurrent), the optical gain (dB) in FIG. 4 was corrected to be half of the electrical gain displayed on the ESA.

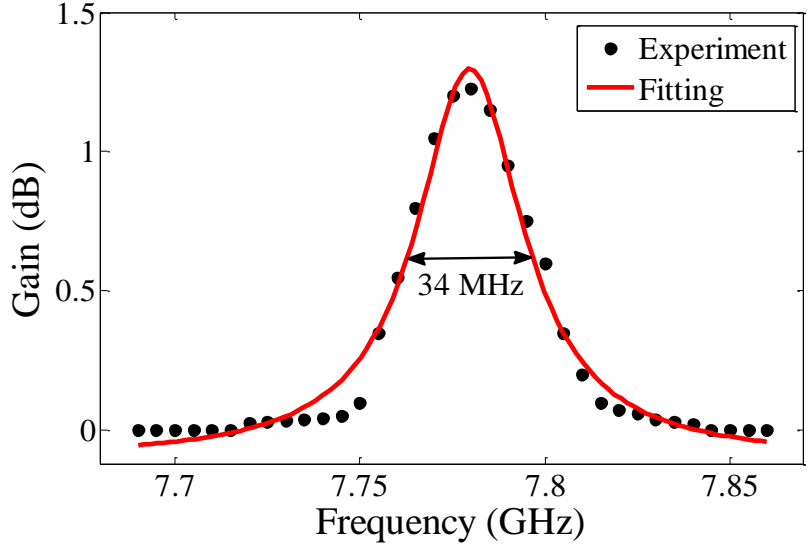

FIG. 4. Measured forward SBS gain spectrum and the corresponding fitting curve for the microring resonator.

Thermal effects in the resonator limited the maximum coupled pump power to about 10 mW. At this power, we measured 1.2 dB Brillouin peak gain at the Brillouin resonance of 7.78 GHz as shown in FIG. 4. The Brillouin resonance matches well with our simulation in FIG. 1 (d). Assuming a Lorentzian fitting function, the Brillouin gain linewidth $\Gamma/2\pi$ is obtained as 34 MHz and the phonon lifetime was calculated as $\tau=1/\Gamma=4.7$ ns and quality factor $Q=\Omega/\Gamma=212$. In the process of SBS, the pump transfers energy to the probe and the gain is the relative change in the intensity of the probe[1].

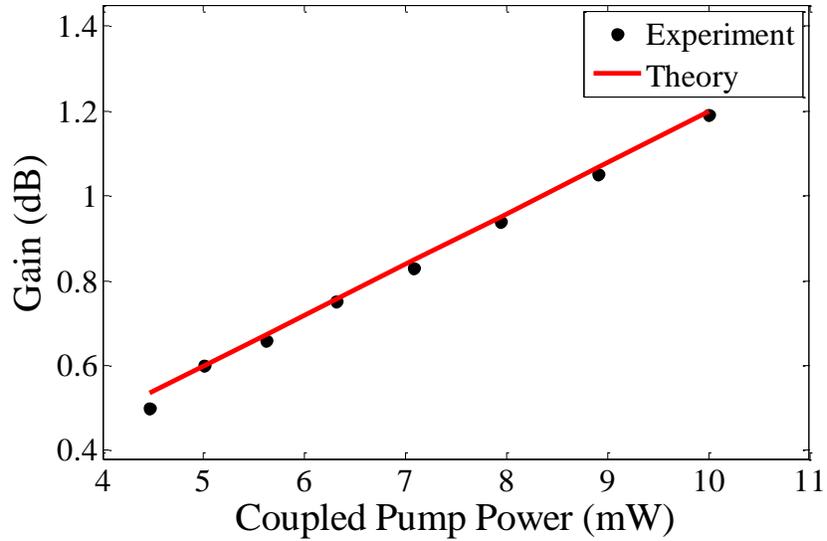

FIG. 5. Measured forward Brillouin gain as a function of coupled pump power.

Keeping the pump wavelength at the resonance and also fixing the probe wavelength, we decreased the coupled pump power and measured the corresponding Brillouin peak gain as shown in FIG. 5. The measured gain is linearly related with the coupled pump power in FIG. 5, which is evidence of negligible nonlinear losses from two photon absorption or free carrier effects at the low pump powers used. Theoretically, the gain can be expressed as

$$G = 10\lg(\exp(G_{SBS} P_p L_{eff})) \qquad (2)$$

where $G_{SBS}$ is the Brillouin gain coefficient, $P_P$ is the coupled pump power, $L_{eff} = (1 - \exp(-\alpha L))/\alpha$ is the effective length with $\alpha$ as the linear loss[16]. For a microring resonator, due to its recirculating nature of the resonant mode, light intensity can be largely enhanced with even short radius. We can equivalently treat the microring resonator as a ring-shaped waveguide but with a long effective interaction length which is determined by

$$L_{eff} = \frac{Q \lambda_{res}}{2\pi n_{eff}} \qquad (3)$$

where Q is the quality factor of the resonator, $n_{eff}$ is the effective refractive index, $\lambda_{res}$ is the resonance wavelength[18]. Thus the effective interaction length is calculated to be 0.62 cm. Using equation (2) and the fitted slope of the relationship between the gain and coupled pump power in FIG. 5, we derive the Brillouin gain coefficient to be 4450 W$^{-1}$ m$^{-1}$.

Since we use the nanoscale structure, there exist electrostrictive and radiation pressure induced forces affecting the generation of SBS[19]. To clarify their contribution to the SBS gain in our structure, we use finite element method to simulate their sign and orientation, respectively. Electrostrictive forces are derived from bulk photoelastic tensors which are largely independent of structure scale and represent the coupling of electromagnetic energy to strain degrees of freedom of a medium through nonzero photoelastic constants[20]. radiation pressure induced forces are derived from Maxwell stress tensors and result from scattering of light at discontinuous dielectric boundaries which are radically enhanced in nanoscale structures[20]. First, we calculate Maxwell stress tensor $T_{ij}$ using equation (4), (5a) and (5b),

$$T_{ij} = \varepsilon_0 \varepsilon \left( E_i E_j - \frac{1}{2}\delta_{ij}|E|^2 \right) + \mu_0 \mu \left( H_i H_j - \frac{1}{2}\delta_{ij}|H|^2 \right) \qquad (4)$$

$$T_{xx} = \varepsilon_0 \varepsilon \left( |E_x|^2 - \frac{1}{2}|E|^2 \right) + \mu_0 \mu \left( |H_x|^2 - \frac{1}{2}|H|^2 \right) \qquad (5a)$$

$$T_{yy} = \varepsilon_0 \varepsilon \left( |E_y|^2 - \frac{1}{2}|E|^2 \right) + \mu_0 \mu \left( |H_y|^2 - \frac{1}{2}|H|^2 \right) \qquad (5b)$$

where $\varepsilon_0$ and $\mu_0$ are the permittivity and permeability in free space, $\varepsilon$ and $\mu$ are the relative permittivity and permeability, $E$ and $H$ are the electric and magnetic field components, respectively. Then we can obtain the radiation pressure induced force densities from $\mathcal{F}_j^{rp} = -\partial_i \sigma_{ij} = \partial_i T_{ij}$. With finite element method, we first simulated horizontal and vertical Maxwell stress tensors as shown in FIG. 6 (a) and (b). Then the horizontal and vertical radiation pressure induced force densities were obtained from the tensors as shown in FIG. 6 (c) and (d), which clearly indicate that those forces exist on the boundaries of our structure.

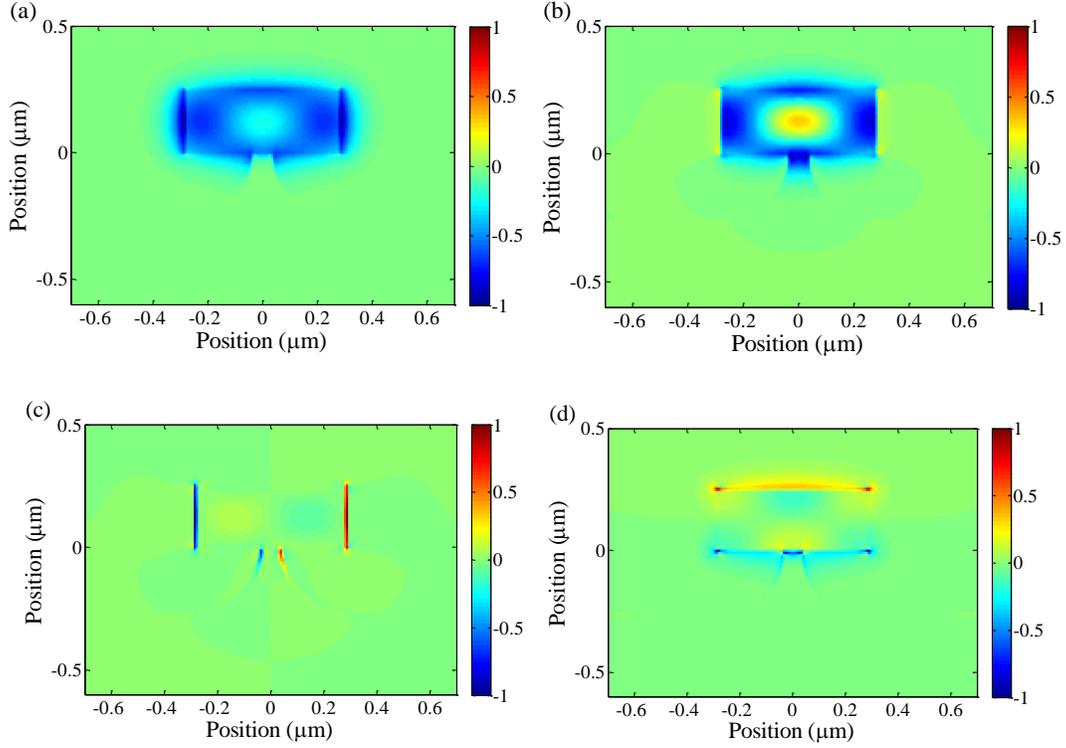

FIG. 6. Simulated Maxwell stress tensor in (a) horizontal and (b) vertical direction; simulated radiation pressure induced force densities of (c) horizontal component and (d) vertical component.

Next, we also calculated electrostrictive force densities through electrostrictive stresses $\sigma_{kl}^{es}$. Electrostrictive stresses are determined with equation (6), (7a) and (7b),

$$\sigma_{kl}^{es} = -\frac{1}{2}\varepsilon_0 n^4 p_{ijkl} E_i E_j \quad (6)$$

$$\sigma_{xx}^{es} = -\frac{1}{2}\varepsilon_0 n^4 \left[p_{11}|E_x|^2 + p_{12}\left(|E_y|^2 + |E_z|^2\right)\right] \quad (7a)$$

$$\sigma_{yy}^{es} = -\frac{1}{2}\varepsilon_0 n^4 [p_{11}|E_y|^2 + p_{12}(|E_x|^2 + |E_z|^2)] \quad (7b)$$

where $n$ is the refractive index, $p_{ijkl}$ is the photoelastic tensor with $p_{11} = -0.09$ and $p_{12} = 0.017$[20]. The electrostrictive force densities are derived from $\mathcal{F}_j^{es} = -\partial_i \sigma_{ij}^{es}$. FIG. 7 (a) and (b) show the simulated electrostrictive stresses in horizontal and vertical direction respectively based on finite element method. Their stress distribution locates around the center of the structure and is close to the distribution of optical mode depicted in FIG. 1 (c) when compared with Maxwell stress tensors shown in FIG. 6 (a) and (b). FIG. 7 (c) and (d) show the horizontal and vertical electrostrictive force densities which clearly illustrates that electrostrictive forces concentrate within the structure and are body forces. FIG. 6 and FIG. 7 indicate that horizontal electrostriction and radiation pressure induced forces dominate in the generation of SBS which interfere constructively to enhance SBS. Since the SBS gain is square proportional to the total photon-phonon overlap, we can benefit about four times larger gain than that results from individual electrostriction or radiation pressure[7].

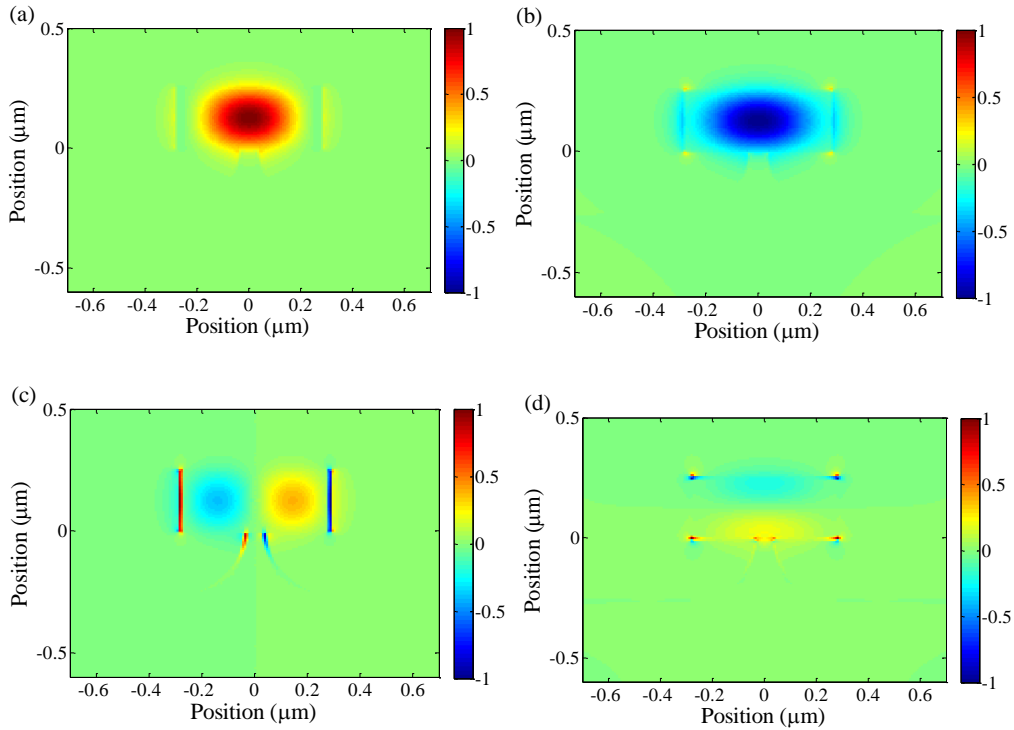

FIG. 7. Simulated electrostrictive stresses in (a) horizontal and (b) vertical direction; simulated electrostrictive force densities of (c) horizontal component and (d) vertical component.

In conclusion, forward SBS has been observed in the silicon microring resonator. With coupled pump power of 10 mW, 1.2 dB Brillouin peak gain is obtained at Brillouin frequency of 7.78 GHz. The Brillouin gain coefficient derived from the measured linear relationship between SBS gain and pump power was 4450 $W^{-1}$ $m^{-1}$. The achieved gain is comparable with previous reports[6,14] but here we used a microring resonator to achieve optical gain in a shorter length and lower coupled pump power.

## ACKNOWLEDGMENTS

This work was fully funded by Hong Kong Research Grant Council (RGC) GRF project 416913.